# Comparison of Data Imputation Techniques and their Impact

Darren Blend and Tshilidzi Marwala
School of Electrical & Information Engineering, University of the Witwatersrand, P/Bag x3, Wits, 2050, South Africa

*Abstract*— **Missing and incomplete information in surveys or databases can be imputed using different statistical and soft-computing techniques. This paper comprehensively compares auto-associative neural networks (NN), neuro-fuzzy (NF) systems and the hybrid combinations the above methods with hot-deck imputation. The tests are conducted on an eight category antenatal survey and also under principal component analysis (PCA) conditions. The neural network outperforms the neuro-fuzzy system for all tests by an average of 5.8%, while the hybrid method is on average 15.9% more accurate yet 50% less computationally efficient than the NN or NF systems acting alone. The global impact assessment of the imputed data is performed by several statistical tests. It is found that although the imputed accuracy is high, the global effect of the imputed data causes the PCA inter-relationships between the dataset to become altered. The standard deviation of the imputed dataset is on average 36.7% lower than the actual dataset which may cause an incorrect interpretation of the results.**

## I. Introduction

THE Acquired Immunodeficiency Syndrome (AIDS) pandemic has already affected more than 36 million people worldwide [1]. Statistics relating to Human Immuno-deficiency Virus (HIV) and AIDS is of paramount importance to national governments and health agencies [2]. The identification of the demographic information regarding infected people can help present the correct preventative measures and bring treatment to the correct locations. The statistical information is most often captured using nationwide surveys which may suffer from vulnerabilities due to badly planned questions leading to missing, or incomplete data [3].

Missing Data is a common phenomenon related to large databases and surveys. The missing data's impact hinders the ability to process data to find conclusive results. Due to the global impact of HIV/AIDS, much effort is in place to gain better insight from statistical information and missing data estimation is of a primary focus.

This report first accounts for the different methods used to help increase the accuracy of data imputation. Three methods are being discussed, namely the benchmark method of an auto-associative (AA) neural network (NN) with a genetic algorithm (GA), an AA neuro-fuzzy (NF) system with a GA and finally a hybrid mixture of a hot-deck (HD) statistical imputation method combined with AANN/NF-GA systems. All methods and results are presented and are also compared under principal component analysis (PCA) dimensionality reduction conditions. The second component of this project is to introduce statistical methods to analyze the global impact of the predicted data and to express the effect data imputation has on the results and conclusions obtained.

## II. Missing Data and Imputation Techniques

Missing data can be caused by missing fields in a database or incorrectly entered information [3]. Depending on the nature of the data and amount of samples available, different imputation methods are available. Fundamentally most imputation techniques can be categorized as either model-based or non-model based [4].

Non-model based techniques include simple omission of records, mean/median substitution and cold or hot-deck imputation which is discussed in section III.C. These methods are consistent, easy to use and preserve the data but limit the variability which model-based imputation techniques provide [5]. These methods are ad-hock and need to be configured and tweaked differently for each application.

Model based techniques were first introduced by Little and Rubin [6] in the 1970's and include AANN and AANF discussed in section III and other regression techniques including maximum likelihood [7]. Model based techniques are more flexible and provide variability in the imputed data outside the statistical standard deviation range.

In addition to the different methods discussed certain procedures enable the efficiency and accuracy of the data imputation to be enhanced [8, 9]. These methods include decision trees/forests, support vector machines and rough set theory. They have the ability to limit the search bounds of the GA or reduce the dimensionality of the dataset through PCA.

Missing data is classified into three distinct classes [6, 7]:
- Missing Completely at Random (MCAR) - The missing value has no dependence on any other variable.
- Missing at Random (MAR) - The missing value is dependent on other variables. The missing data pattern can thus be traced by viewing the other variables.
- Missing Not at Random (MNAT) – The missing value depends on other missing values and thus missing data imputation cannot be performed from the existing data.

For this project we need to assume that the data is MAR, which implies that the missing values are deducible in some complex manner from the remaining data [7].

## III. Background of Proposed methods

### A. Auto-Associative Neural Networks

The benchmark method to which all the subsequent methods will be compared is an auto-associative neural network with genetic algorithm (AANN-GA). This method was first presented by Abdella and Marwala in 2005 [5] and has since been the topic of numerous academic papers.

*1) Neural Networks*

A neural network is a soft-computing technique based on the physiological working of the biological brain [10]. Neural network architecture consists of separate layers of nodes or neurons, usually an input, hidden and output node layer. Neural networks learn from training and hence experience, rather than from being mathematically programmed. This ability allows for neural networks to work on complex non-linear systems as there are no rules or formula that govern their operation. Each time the Neural Network is trained, internal changes are made to the pathways between the nodes [11]. There are different NN architectures including multi-layer perceptron (MLP) and radial basis function (RBF). A typical NN learning algorithm uses back-propagation which compares the predicted to the desired output and propagates the error found by adjusting the nodal weights. A non-linear optimization technique such as scaled gradient decent, allows for the non-linearity of the dataset to be captured in the model.

*2) Auto-Associative*

An auto-associative or auto-encoder neural network is a specific configuration of a NN which is trained to recall the input dataset [7]. The inputs and outputs of the neural network are made identical. The AANN can map both linear and non-linear relationships between the inputs dataset. Fig. 1 shows a typical butterfly diagram of the AANN showing the smaller number of hidden nodes to input/output nodes creating a bottleneck situation.

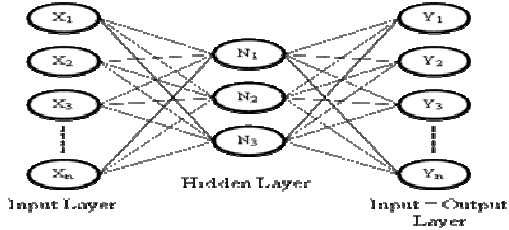

Fig.1: Diagrammatic representation of an AANN.

*3) Genetic Algorithm*

A GA is a "survival of the fittest" search algorithm which is designed based on the Darwinian explanation of evolution [12]. Making use of reproduction between members of a population, offspring in the form of chromosomes can be formed. The governing laws for reproduction follow a principle outlined by the fitness function. This states that chromosomes with a more optimized solution will breed more frequently than others. In addition to the fitness function, offspring are subject to two other phenomenon, namely recombination (cross-over) and mutation. GA through the use of these phenomenon allow for the global maximum to be found and prevent premature convergence or local maximums which is a problem facing other search algorithms [13].

*4) Method Overview*

The methodology outlined by the AANN-GA, is presented in Fig. 2. The AANN is trained on a complete dataset. The GA allows for the estimation of the unknown missing inputs and the accuracy of the imputation is based on the minimization of the AANN error through the GA fitness function [7].

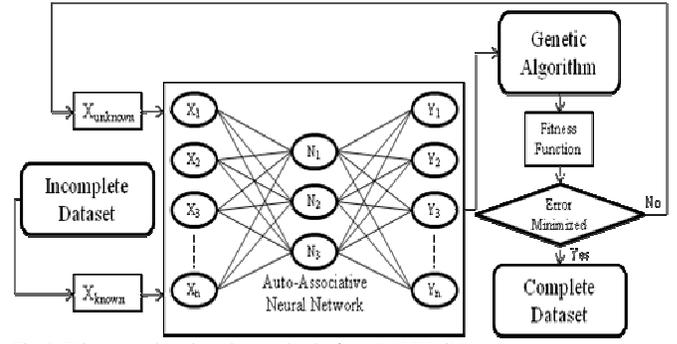

Fig.2: Diagram showing the method of an AANN-GA.

The GA fitness function is used to minimize the error of the AANN. The error function is shown in equation 1.

$$e = \left( \begin{Bmatrix} X_k \\ X_u \end{Bmatrix} - f\left( \begin{Bmatrix} X_k \\ X_u \end{Bmatrix}, \vec{W} \right) \right)^2 \quad (1)$$

$X_k$ is the known inputs, while $X_u$ is the unknown or missing data values. The function $f$ represents the NN or NF presented in section III.B with the network weights shown by $W$.

*B. Neuro-Fuzzy Networks*

NF systems incorporate the advantages of both the NN and fuzzy-logic (FL) soft-computing techniques [14]. The accurate input/output quantitative relationships obtained from the NN and the intrinsic qualitative knowledge of the system composition obtained from the FL component allow for a fully comprehensive system model [15].

*1) Fuzzy Inference System*

A fuzzy inference system (FIS) is a class of feed forward systems based on linguistic rules and abstractions [15]. Fuzzy logic was first introduced by Lofti Zadeh in 1965 and allows for variables to have a varying level of truth and need not be discrete [16]. By using expert human rules and membership functions (MF), an abstraction from the quantitative complexity of non-linear systems can be achieved. FIS are not trained on known inputs and can thus be used with limited data examples. FIS can be of either Sugeno or Mamdani architectures. The main difference being that Mamdani systems use MF for both the inputs and outputs while Sugeno uses MF for only the input layer [15].

*2) Adaptive Neuro-Fuzzy Inference System*
  *a) Architecture*

There are currently numerous different NF architectures available [17]. Adaptive Neuro-Fuzzy Inference System (ANFIS) is a FIS and allows the membership functions and rule weightings to be adaptively altered through the use of a trained NN. Thus the combination of quantitative NN and qualitative FL systems are combined. ANFIS contains both non-adaptive and adaptive nodes and is a 5 layered system shown in Fig. 3 [18]. ANFIS is a of a Sugeno FIS architecture. Layer 1 represents the membership functions that can adaptively be altered. Layer 2 and 3 are fixed nodes which perform multiplication and normalization of data respectively. Layer 4 represents the fuzzy-rules weightings which have the ability to be altered. The final layer 5 represents a summation which calculates the appropriate output.

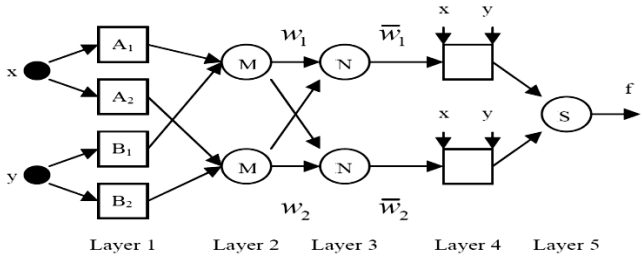

Fig 3: A typical ANFIS Neuro-Fuzzy architecture [18].

*b) Subtractive Clustering*

Because of the high dimensionality and complex non-linear relationships between the input dataset, fuzzy rules cannot be determined from expert knowledge alone. Clustering techniques are capable of identifying natural groupings within data and organize data into distinct classes based on the determined similarities [15]. Fuzzy rules can then be extracted from the distinct data classes. Subtractive clustering uses cluster centers with finite radii surrounding them to generate groups within data. This technique takes longer to compute but produces a more optimized system with fewer rules. Due to the high dimensionality of the dataset, subtractive clustering allows data points to be used as the candidates for cluster centers, as approximations to the accurate grid points as used in grid partitioning [19].

*3) Principal Component Analysis*
*a) Background*

Due to the exponential increase in FL rules with increase dataset dimensionality, the computational time for training a NF system becomes unfeasible for high dimensionality data [15]. The logical solution to this problem is too either exclude inputs from the dataset or use an encoding system to approximate the dataset with a lower dimensionality. PCA is a statistical technique which finds patterns and commonality in high dimensionality data [20]. By selecting the most important principal components of the data, the dataset can be encoding to a lower dimensionality which can be used to for training.

*b) Algorithm*

The PCA algorithm to compress the dataset is outlined [21]:
1. The mean (*OriginalMean*) of each dimension is subtracted from each value. (*DataAdjust*)
2. The covariance matrix of the dataset is determined.
3. The eigenvalues and eigenvectors of the dataset are determined.
4. The eigenvalues are ordered in descending order of magnitude. The higher eigenvalues correspond to the more important principal components.
5. The degree of compression is dependent on the amount of principal eigenvectors chosen. These eigenvectors will form the feature vector which is effectively the PCA network and used to compress and uncompress the dataset. (*FeatureVector*)

Equations 2 and 3 [21] show the required formula for data compression (*FinalData*) and hence dimensionality reduction and for the reverse procedure of obtaining the original dataset (*OriginalData*), where $^T$ represents the transpose matrix.

$$FinalData = FeatureVector \times DataAdjust \quad (2)$$

$$OriginalData = (FeatureVector^T \times FinalData) + OriginalMean \quad (3)$$

As with most encoding systems, there is data loss when reconstructing the original data matrix and this error is proportional to the magnitude of the dimensionality reduction used in the compression [21].

*4) Method Overview*

The NN or NF system is trained to map the original high dimensionality inputs to the *FinalData*, reduced dimensionality dataset. The genetic algorithm thus performs a PCA compression using the same *FeatureVector* network and an error function can be computed as shown in equation 4.

$$e = \left( f_{PCA}\left( \begin{Bmatrix} X_k \\ X_u \end{Bmatrix}, \vec{U}_K \right) - f_{NF}\left( \begin{Bmatrix} X_k \\ X_u \end{Bmatrix}, \vec{W} \right) \right)^2 \quad (4)$$

A diagram representing the complete AANN/AANF-PCA-GA system is presented in Fig. 4.

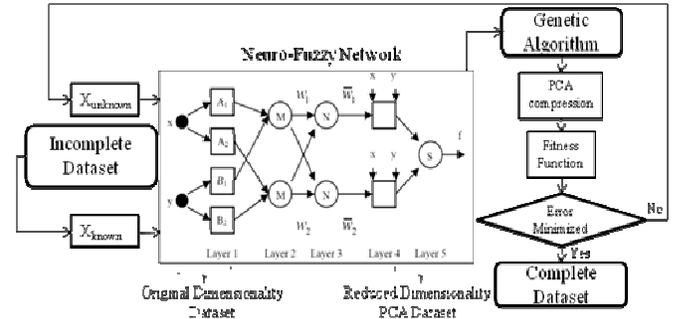

Fig. 4: Diagram showing the method of the AANN/AANF-PCA-GA.

*C. Hybrid System*

The accuracy of many model-based imputation methods is greatly improved by limiting the bounds of the GA search space [8]. This improves both the efficiency of computation and can provide more consistent results.

*1) Hot-Deck Imputation*

HD imputation is a non-model based statistical method [3]. HD imputation is the most widely used data imputation method. There are numerous different HD configurations but all revolve around locating similar dataset matches. The mean of the similar matches is then computed. By adding a range to the calculated mean, the upper and lower bounds of the GA can be determined. This method introduces advantages of simplicity of design but is computationally inefficient [5].

*2) Method Overview*

A common problem with non-model based systems is the lack of variability in the dataset [5]. This problem is overcome with two original procedures, firstly instead of selecting a single most similar matching case from the original dataset, a collection of a minimum of six similar cases are found and the mean calculated from them all. The similar matches are determined by having a dynamically adjusting error function to compute the degree of similarity. Secondly to maintain variability in the dataset, the upper and lower GA bounds are placed one standard deviation apart from the calculated mean. This affords the GA the freedom to find the best imputed value within the statistically determine space. Fig. 5 presents a flowchart of the hybrid combination of HD with an AANF.

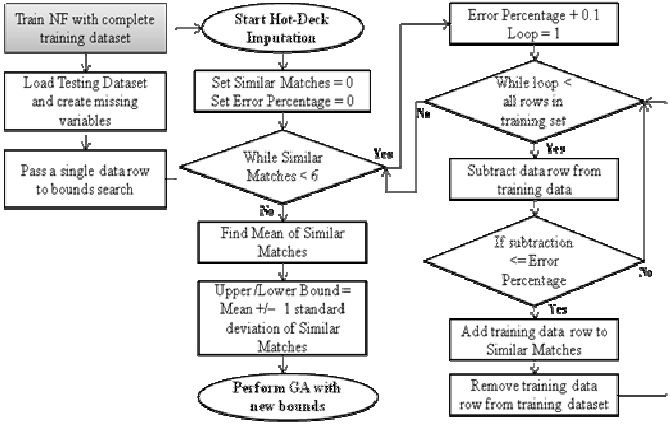

Fig. 5: Flowchart of a hybrid algorithm combining an NF with hot-deck.

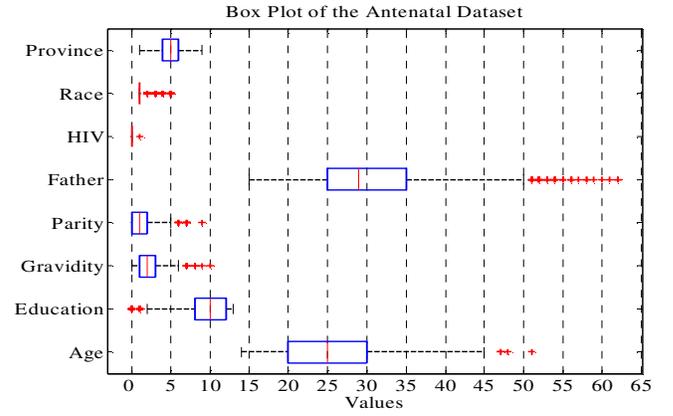

Fig. 6: Box Plot of the dataset's categories and their range, mean and outliers.

## IV. PRENATAL AIDS DATASET AND PREPROCESSING

The dataset used for this project is from an antenatal clinic survey conducted in 2001. There are 12089 samples available which is sufficient to split the dataset into three groups namely training, validation and testing. The groups are split with a 2:1:1 ratio with twice as many samples in the training dataset. The dataset consists of 8 categories and are shown in the box plot in Fig. 6 showing the range, mean and statistical outliers.

Preprocessing of the dataset is crucial before any of the imputation methods can be performed. Qualitative variables like race and location need to be binary encoded in order to prevent incorrect weighting issues in the MLP NN [1]. Outlier detection is performed primarily through visual plotting of the data and secondly through logical statement such as parity always being smaller than gravidity. String fields in the dataset are encoded to integer values using a lookup table.

The entire dataset is normalized to fall between 0 and 1. This is done to prevent high magnitude values from having a biased weighting in the different algorithms. When splitting the dataset into the different groups, randomization of the dataset is constantly performed and repeated iterations are performed to try match the different dataset mean values such that the PCA is able to perform accurately.

## V. EXPERIMENTS AND RESULTS

### A. Background

There are numerous different data imputation methods and papers describing different aspects of data prediction. The goal of this report is to account for missing data accuracy in the fields of age, education and HIV status. These categories are selected out of the entire dataset in order to obtain information regarding the highest level of education which most women complete. This information is very useful in directing sex education to the appropriate education level. Government and educational agencies can use this information and direct the preventative AIDS courses at the appropriate level thereby targeting the largest audience and cause the most effective prevention. This report first presents the results obtained for the different imputation methods and records the accuracy of prediction. The second component presented in section VI is to utilize the new predicted data and show statistically how the global impact of the imputed data differs from the actual data.

### B. Classification of Imputation Error

To determine the accuracy of the imputation technique two different error quantifications can be used. The first is the root mean square error (RMSE) [8]. This error is however only meaningful when it is with respect to an equivalent error.

The second error method finds the percent of predicted values which fall into different classes based on the prediction within a given number of years or units, such as:

- Age: Within 2, 4 and 6 years
- Education Level: within 1, 2 and 3 grades
- HIV status: According to binary classification following formula 5.

$$Accuracy_{HIV} = \frac{No.\,True\,Positives\,+\,No.\,True\,Negatives}{Total\,No.\,of\,Data} \quad (5)$$

### C. Experimental Results

The experimental tests conducted are designed to give an indication of the accuracy of the imputation ability for the different methods discussed in the previous sections.

The MLP AANN is configured with 11 hidden nodes and 250 training cycles and uses the *netlab* toolbox. The NF uses ANFIS with 30 training cycles and subtractive clustering rule extraction with a radius of 0.3. The GA uses the *gaot* toolbox with a population size of 50 and 20 generations. The results for the data imputation methods, as well as the hybrid combination of the hot-deck imputation are shown in Table I.

From Table I it can be seen that the accuracy of both the NN-GA and NF-GA methods is greatly improved when combining the hot-deck imputation algorithm. The compromise for the increased prediction accuracy is a decrease in computational efficiency by 50%. The NN-GA has an average improvement of 10.6% and 15.1% for the 1 and 3 missing columns respectively, while the NF-GA has an average improvement of 22.1% and 36.2% for the 1 and 3 missing columns respectively when using the hot-deck imputation. The results also conclude that the NN based methods are far superior to the NF techniques for both the normal conditions and the hybrid hot-deck combination in terms of accuracy and computation efficiency.

Table I: Percentage accuracy of the different data imputation methods.

| Missing Columns | Age | Education | HIV | Age/Education/HIV | | |
|---|---|---|---|---|---|---|
| # of Missing Columns | 1 | 1 | 1 | 3 | | |
| Class | 2 year | 1 grade | HIV | 2 year | 1 grade | HIV |
| | 4 year | 2 grade | | 4 year | 2 grade | |
| | 6 year | 3 grade | | 6 year | 3 grade | |
| NN-GA (%) | 49.0 | 40.0 | 63.6 | 42.0 | 35.7 | 62.4 |
| | 74.3 | 63.5 | | 66.5 | 55.7 | |
| | 87.8 | 78.9 | | 84.1 | 71.7 | |
| NN-GA-with Hot Deck Imputation (%) | 61.5 | 53.8 | 74.1 | 55.2 | 51.3 | 77.9 |
| | 85.2 | 75.4 | | 83.2 | 74.4 | |
| | 94.3 | 87.1 | | 93.3 | 88.2 | |
| NF-GA (%) | 42.6 | 43.7 | 40 | 22.3 | 24.7 | 38.3 |
| | 59.6 | 56.7 | | 39.0 | 41.3 | |
| | 69.4 | 69.0 | | 50.3 | 54 | |
| NF-GA-with Hot Deck Imputation (%) | 57.0 | 55.3 | 72.7 | 51.0 | 50.7 | 77.6 |
| | 85.0 | 70.7 | | 82.3 | 72.3 | |
| | 92.7 | 81.0 | | 94.3 | 89.0 | |

The NN, NF and both hybrid hot-deck imputation methods are also compared under PCA conditions. The PCA is chosen to compress the data by 2 dimensions to produce 11 inputs form the original 13. Although the data compression only chooses the most principal components of the dataset and discards the rest, the results obtained and which are tabulated in Table II, are comparable to the results presented in Table I. The NN based methods again outperform the NF methods by an average of 4.8% and 13.6% for 1 and 3 missing columns respectively and by 2.9% and 2% for 1 and 3 missing columns respectively for the hybrid hot-deck imputation methods. The full dataset of Table I, for both NN and NF compared to the PCA methods in Table II, are on average 1.2% and 5.7% for 1 and 3 missing columns respectively less accurate and for the hybrid hot-deck imputation method by 4.0% and 1.4% for 1 and 3 missing columns respectively more accurate.

Table II: Percentage accuracy of the different data imputation PCA methods.

| Missing Columns | Age | Education | HIV | Age/Education/HIV | | |
|---|---|---|---|---|---|---|
| # of Missing Columns | 1 | 1 | 1 | 3 | | |
| Class | 2 year | 1 grade | HIV | 2 year | 1 grade | HIV |
| | 4 year | 2 grade | | 4 year | 2 grade | |
| | 6 year | 3 grade | | 6 year | 3 grade | |
| NN-PCA-GA (%) | 57.2 | 31.5 | 65.4 | 55.1 | 31.7 | 64.7 |
| | 82.8 | 41.6 | | 79.3 | 46.3 | |
| | 91.8 | 53.5 | | 91.1 | 59.8 | |
| NN-PCA-GA-with Hot Deck Imputation (%) | 58.4 | 48.3 | 70.5 | 60.6 | 48.4 | 77.2 |
| | 82.4 | 68.9 | | 82.4 | 68.3 | |
| | 90.7 | 82.3 | | 93.4 | 83.1 | |
| NF-PCA-GA (%) | 45.0 | 39.3 | 55.6 | 27.8 | 30.8 | 58.2 |
| | 63.7 | 55.0 | | 43.4 | 46.2 | |
| | 73.7 | 68.2 | | 53.4 | 59.0 | |
| NF-PCA-GA-with Hot Deck Imputation (%) | 52.6 | 51.6 | 67.4 | 56.2 | 47.8 | 76.6 |
| | 77.8 | 72.4 | | 80.4 | 70.8 | |
| | 87.4 | 81.2 | | 90.4 | 83 | |

## VI. STATISTICAL IMPACT OF THE ESTIMATED DATA

### A. Background of Statistical Methods

Based on the results obtained in section V, the most novel method which also produced the best results is the NN-GA with hot-deck imputation. This method will be used to perform the statistical impact of the missing data.

This section deals with the impact which the new predicted values have on the observations which can be concluded from the purpose of the data imputation procedure. For this project the focus of the data imputation is placed on the ability to predict the age, education level and HIV status such that recommendations can be made as to which level of schooling, sex education preventative classes should be conducted.

To gain the global statistical effect of the data imputation different statistical methods are used and each is described in detail below. The different tests help conclude whether the proposed methods of data imputation improve or degrade the information abstracted from the antenatal survey.

### B. Statistical Methods, Results and Observations

#### 1) Mean and Median

The arithmetic mean is the average or expected value of a dataset, while median is the second quartile or the middle point in a dataset [22]. Due to the unequal distribution of the dataset, the mean and median are not identical. This results in the mean of the dataset being biased by high magnitude outliers. The median is thus a better indication of the normal age, education level and HIV status for the sample dataset. The mean and median values are tabulated in Table III.

Table III: Actual and Imputed Mean and Median of the Dataset

| | Age | Education | HIV Status |
|---|---|---|---|
| Actual Mean | 25.62 | 9.34 | 0.21 |
| Imputed Mean | 25.27 | 9.66 | 0.06 |
| Actual Median | 25.00 | 10.00 | 0.00 |
| Imputed Median | 24.00 | 10.00 | 0.00 |
| Actual Std Dev | 6.58 | 2.90 | 0.40 |
| Imputed Std Dev | 5.22 | 1.46 | 0.24 |
| Correlation Index | 0.84 | 0.38 | 0.10 |

#### 2) Standard Deviation and Probability Density Function

The standard deviation represents the measure of the dispersion or spread of the dataset [22]. This value informs how closely all the values of the dataset are clustered around the mean of the dataset. The standard deviation is able to show what the probabilities are of obtaining a data point near the extremities of the dataset range. The standard deviations are shown in Table III. The discrepancies between the values give an indication to the degree of spread. The probability density function (PDF) is a histogram of the frequency of occurrence of a dataset and is shown in Fig. 7 for the age data [23]. It can be seen that the imputed data fall into a narrower spread with most values about the data mean. This is as a result of the hot-deck imputation having a lack of variability.

#### 3) Correlation Coefficient and Quantile Plot

The correlation coefficient gives an index to the strength of the relationship between two variables [22]. It is a value ranging from -1 to 1 with 0 being completely unrelated. Table III shows the correlation between the actual and imputed data variables. It can be seen that the age column has the highest

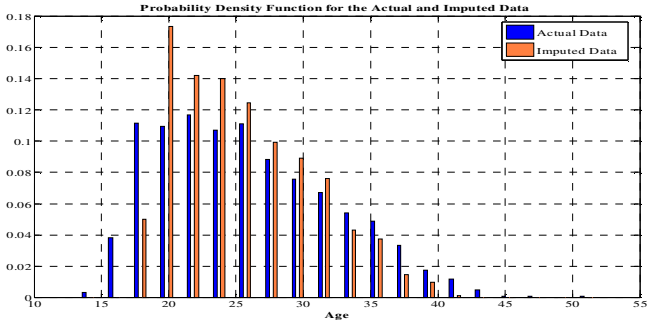
Fig. 7: PDF plot for the actual (blue) and imputed (yellow) age data.

correlation index, indicating it is the most accurate imputation.

If the quantile-quantile plot shown in Fig. 8 is linear it shows if the actual and imputed datasets are from the same distribution [23]. This is important to ascertain whether the imputed data is a fully comprehensive representative of the actual dataset.

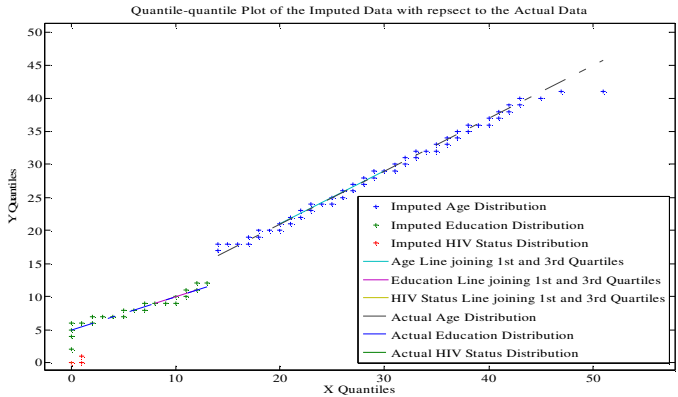
Fig. 8: Quantile plot of linear relationships representing common distributions

### 4) PCA Using Factor Analysis

PCA described in section III.B is used to find the important principal components in a dataset such that dimensionality reduction can be performed through a compression algorithm. Using the same principles, the linear relationships between dataset variables can be visualized. A comparison between the dataset variables inter-relationships is shown in Fig. 9 for the actual and imputed datasets for the first three principal components. It can be seen that the inter-relationships for these principle components are different between the actual

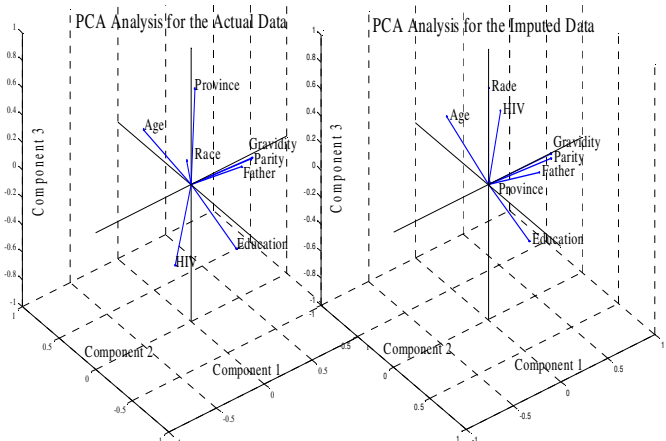
Fig.9: PCA analysis for the first 3 components for the actual and imputed data

and imputed datasets. Fig. 10 shows the weighting of the principal components for the imputed and actual datasets. It can be seen that the first three principal components amount for as much as 75% of the complete data information [23]. Comparing the percentage composition of the principal components for the imputed and actual dataset, a discrepancy in the results can be observed. The results differ especially for the first principal component which is 6% higher for the imputed data. The principal component information provides an indication to the global impact that imputation of data has not only to the missing data columns, but to the dynamics of the entire dataset.

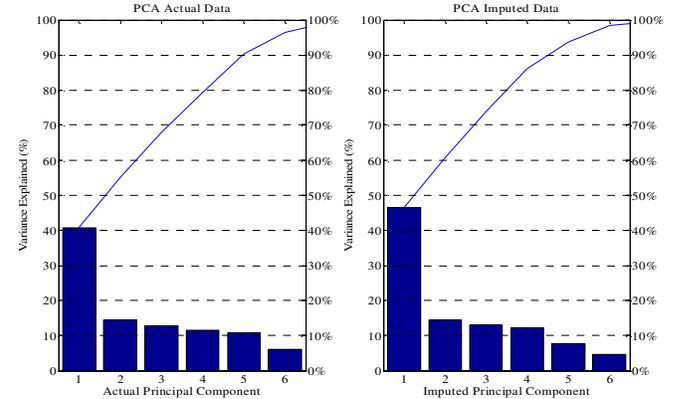
Fig. 10. Percentage composition of the different principal component values.

### 5) Data Classification

The final test performed to analyze the global impact of the imputed data is the use of the predicted data for classification purposes. Using the predicted data, the HIV status will be determined and the accuracy of the prediction compared to the classification with the actual dataset. This test is able to give an indication of the degree to which the data imputation techniques presented in the previous sections is affected by the imputed data. NN and NF systems are trained with the original training dataset to predict the HIV status. The actual and imputed testing dataset are then passed through the network to classify the HIV status. Table IV presents the results obtained for the two datasets.

Table IV: Classification percentage accuracy for the actual and imputed data

| Method | Dataset | HIV Status (%) |
|---|---|---|
| NN | Actual | 68.6 |
| | Imputed | 66.5 |
| NF | Actual | 60.4 |
| | Imputed | 59.0 |

The results obtained show that the discrepancy between the classification accuracy when using the actual dataset opposed to the imputed dataset for the age and education categories is 2.1% for the NN and 1.4% for the NF system. This accuracy difference is insignificant. The conclusion drawn from this test indicates that the accuracy of imputation for two missing columns is sufficient to replace the original dataset when predicting the HIV status of an individual.

### VII. DISCUSSION OF RESULTS AND FUTURE WORK

The first task of obtaining an accurate imputation method is successfully implemented using the hybrid hot-deck imputation method. The success is attributed to the complementary combination of a consistent non-model based statistical method with a more flexible model-based network. By introducing a range of a standard deviation for the upper and lower GA bounds for the hot-deck imputation technique, the hybrid method is generic and is able to be used on different datasets with varying data ranges.

The statistical analysis shows how the global impact of the imputation is able to cause variations in the results. By utilizing the median level of education and subtracting the standard deviation value, the actual data suggests that sexual education should be performed at the 7$^{th}$ grade such that one standard deviation or 68% of the sample population will still be in school [22]. The imputation data suggests the sexual education must take place half way through the 8$^{th}$ grade such that the same percentage of the sample will have access to the sexual education.

The correlation coefficient index is highest for the age data and lower for the education and HIV status indicating a lack of consistancy in results. The quantile plot however recognises that the actual and imputed data are from the same distribution which shows that the imputed data is a fully comprehnsive representation of the actual data. The PCA analysis of the data is useful for showing how much of an impact the imputed data is able to cause not only to the missing data variables but to the entire datasets inter-relationships.

Future work for this project can make use of a comparison between GA and particle swarm optimization to find a better suited search algorithm [13] and investigating Monte Carlo - Markov chain simulations to optimize the errors and alogirthms used [24].

## VIII. CONCLUSION

A novel hybrid combination of a non-model based statistical HD imputation method with a model-based NN or NF is able to improve the accuracy of imputation by up to 36.2% for NF and 15.1% for NN. In all the results the NN outperform the NF system both in terms of accuracy and computational efficiency. The PCA with a two dimension compression performed with an average of 0.6% better accuracy against the full dataset and is thus a preferred method based on its superior computational efficiency.

Although the accuracy of imputation is found to be very high, the global effect of the imputed data is shown to be statistically different from the actual data. By plotting a PDF histogram distribution, it is found that due to the HD imputation the imputed data has a lack of variability in its data causing a lower standard deviation and less spread of data about the mean. The PCA analysis of the data shows how the complex inter-relationships between the data categories are skewed due to the imputation of data. However, when using the imputed data to replace categories in the original dataset, the classification accuracy of the HIV status is almost equivalent to the actual dataset's classification ability.

The final conclusion for the level of schooling at which sexual education should be implemented shows that for one standard deviation of 68% of the sample population the actual data suggest the 7$^{th}$ grade, while the imputed suggests half way through the 8$^{th}$ grade. This discrepancy can negatively affect HIV prevention.